\documentstyle[12pt]{article}
\begin{document}
\def\y{\rule{0.5pt}{7pt}}
\def\yy{\rule[-1mm]{0.6pt}{12pt}}
\begin{center}
   {\large Harmonic Map Formulation of }  \\[5mm]
   {\large Colliding Electrovac Plane Waves$^*$}\\[35mm]
   {\large Yavuz Nutku } \\[2mm]
     Department of Mathematics \\ Bilkent University \\
     06533 Ankara, Turkey \\[35mm]
\end{center}
The formulation of the Einstein field equations admitting two Killing vectors
in terms of harmonic mappings of Riemannian manifolds is a subject in which
Charlie Misner has played a pioneering role. We shall consider the hyperbolic
case of the Einstein-Maxwell equations admitting two hypersurface orthogonal
Killing vectors which physically describes the interaction of two electrovac
plane waves. Following Penrose's discussion of the Cauchy problem we shall
present the initial data appropriate to this collision problem. We shall
also present three different ways in which the Einstein-Maxwell equations
for colliding plane wave spacetimes can be recognized as a harmonic map.
The goal is to cast the Einstein-Maxwell equations into a form adopted to
the initial data for colliding impulsive gravitational and electromagnetic
shock waves in such a way that a simple harmonic map will directly yield
the metric and the Maxwell potential 1-form of physical interest.\\[5mm]
$^*${\it for Charles W. Misner on his 60$^{th}$ birthday}

\section{Introduction}

Charlie Misner was the first to recognize that the subject of harmonic 
mappings of Riemannian manifolds finds an important application in general 
relativity. In a pioneering paper with Richard Matzner \cite{mm} he found
that stationary, axially symmetric Einstein field equations can be formulated 
as a harmonic map. Eells and Sampson's theory of harmonic mappings of 
Riemannian manifolds  \cite{es} provides a geometrical framework for thinking
of a set of {\it pde}'s, in the same spirit as ``mini-superspace" that
Charlie was to introduce \cite{w} for {\it ode} Einstein equations a little
later. The subsequent developement of the subject of space-times admitting
two Killing vectors, that eventually led to its recognition as a completely
integrable system \cite{maison} - \cite{neuk} has employed another
formulation of the stationary, axi-symmetric field equations due to
Ernst \cite{e} which is equivalent to that of Matzner and Misner.

Charlie's later work on harmonic maps \cite{m} encompasses a scope much 
broader than this specific problem and its power and elegance is bound to 
make a major impact on theoretical physics.

I was privileged to be in contact with Charlie's ideas at that time
and worked on the two Killing vector problem \cite{n}, \cite{en}.
It was the hyperbolic version of gravitational fields admitting two Killing
vectors that attracted my attention. This is the problem of colliding 
impulsive plane gravitational waves for which Khan and Penrose had presented 
a famous solution \cite{kp}. My work finally led to the exact solution for 
colliding impulsive plane gravitational waves with non-collinear polarizations 
\cite{nh} which is physically the most general solution of this type.
It turned out that the Matzner-Misner formalism was the one more readily 
amenable to the hyperbolic problem, even though it was originally intended 
for the elliptic case, whereas the Ernst formulation fitted the elliptic 
problem, {\it ie} the exterior field of rotating stars, best. The relationship 
between these two formalisms is given by a Neugebauer-Kramer 
involution \cite{nk}.

A few years after the solution \cite{nh} appeared, there was a remarkable 
avalanche of papers on colliding plane gravitational waves. There were 
important papers examining the singularity structure of spacetimes resulting 
from the collision of gravitational waves \cite{t} - \cite{y}. However,
there was also a mass of new exact solutions which are all essentially devoid 
of any physical interest because their authors had chosen not to solve the 
Cauchy problem with the initial data appropriate to generic plane waves, but 
rather they started with a ``solution" and {\it derived (!)} the initial data.
This type of derived initial data for the collision problem describes some 
very peculiar plane waves indeed. An inordinately large number of such
references can be found in \cite{g}.

Nevertheless, physically interesting colliding wave problems are
still open and waiting for an exact solution! Remarkably enough,
the interaction of plane impulsive gravitational and
electromagnetic shock waves is in this category.
We have the Khan-Penrose and Bell-Szekeres  \cite{bs} solutions
describing the interaction of either two impulsive gravitational,
or two electromagnetic shock waves alone
and also the solution of Griffiths \cite{gev}
for the interaction of an impulsive gravitational wave with
an electromagnetic shock wave. But the generic case where
we must consider the collision of both type of waves is missing
even in the case of collinear polarization.
The important open problem here is the construction of an exact solution
of the Einstein-Maxwell equations that reduces to all, the
Khan-Penrose, Bell-Szekeres  and Griffiths solutions.
There are various unsatisfactory treatments of this
problem in the literature \cite{cx}, \cite{gx}.
I shall give its harmonic map formulation.

\section{Initial Data}

The problem of colliding plane gravitational waves was proposed and in
essence solved by Penrose \cite{p1} in 1965 even though most people
writing on this subject do not seem to be familiar with it.
We shall use Penrose's formulation  of the Cauchy problem \cite{goldie},
\cite{bible} to discuss the interaction of
two plane waves, every one of which will consist of a superposition
of an impulsive gravitational wave and an electromagnetic shock wave.
The interaction will be determined by an integration of the Einstein-Maxwell
equations with initial data defined on a pair
of intersecting null characteristics.
The initial values of the fields will be those appropriate to a plane wave
which is given by the Brinkman metric \cite{b}, \cite{ek}
\begin{equation}
  ds^{2} = 2\,d\,u' \;d\,v' \; - \;d\,x'^{\;2}\; - \;d\,y'^{\;2}
+ 2\, H(v',x',y')\, d\,v'^{\;2}                                \label{pw}
\end{equation}
and the superposition of an impulsive gravitational wave and an 
electromagnetic shock wave, with amplitudes proportional to
$a, b$ respectively, is obtained for
\begin{equation}
 H = \frac{a}{2} \left( y'^{\;2} -  x'^{\;2} \right) \, \delta (v')
    - \frac{b^2}{2} \left( x'^{\;2} +  y'^{\;2} \right) \, \theta (v')
\label{hdata}
\end{equation}
where $ \delta $ is the Dirac delta-function and $\theta$ is the
Heaviside unit step-function.

The Brinkman coordinate system employed in eq.(\ref{pw}) is useful
because the superposition of waves travelling in the same direction
is obtained simply by addition.
However, the Brinkman coordinates are
not suitable for the collision problem because of the explicit
dependence of the metric on $x' , y' $.
For this purpose we must transform to the Rosen form
where the metric coefficients will depend on $v$  alone.
This is accomplished by the Khan-Penrose transformation
\begin{equation} \begin{array}{lcl}
v' & = &  v \,,\\
u' & = & u \, + \, \frac{1}{2} x^2 \, F \, F_v \,
 + \, \frac{1}{2} y^2 \, G \, G_v \, \,,\\
x' & = & x \, F \,, \\
y' & = & y \, G \,,
\end{array}  \label{kptr}
\end{equation}
which results in
\begin{equation}
  ds^{2} = 2\, d\,u \;d\,v \; - \;F^2 \,d\,x^2\; - \;G^2 \,d\,y^2 \,,
\label{kprosen}
\end{equation}
provided
\begin{equation} \begin{array}{lcl}
F_{vv} & = &\left[
\; - a \, \delta (v)  - b \, \theta (v) \;  \right] \; F  \,, \\
G_{vv} & = &\left[ \;  - a \, \delta (v)
  + b \, \theta (v) \;  \right] \; G \, .
\end{array}  \label{kpeq}
\end{equation}
These are linear, distribution-valued ordinary differential equations
which can be solved using the Laplace transform
\begin{equation}
{\cal F}(s) = \int_0^{\infty} e^{s v} F (v) dv
\label{laplace}
\end{equation}
and from eq.(\ref{kpeq}) we find
\begin{equation}
{\cal F}(s) = \frac{1}{s^2 + b^2 } \left[ ( s - a ) F(0) + F_v(0) \right]
\end{equation}
where of $F(0) \,, \, F_v (0) $ are initial values.
They are obtained from the continuity of the metric
and its first derivatives across $v=0$ which requires
$F(0)=1$, $F_v(0)=0$. In this case inverting the Laplace
transform we get
\begin{equation}
 F = cos ( b v \theta(v) ) -
 \frac{a}{b} \, sin ( b v \theta (v) )     \label{iv}
\end{equation}
and the result for $G$ is obtained by letting $a \rightarrow - a$
in eq.(\ref{iv}) as indicated by eqs.(\ref{kpeq}).

In Rosen coordinates the general form of the metric that admits two
hypersurface orthogonal Killing vectors is given by
\begin{equation}
  ds^{2} = 2\,e^{-M} \; d\,u \;d\,v \;
-\; e^{-U}\; \left( \,e^V\,d\,x^2\; + \;e^{-V}\,d\,y^2 \, \right)
\label{rosen}
\end{equation}
where $U, V, M$ depend on only $u,v$
and comparison with eqs.(\ref{kprosen}) and (\ref{iv}) shows that
for the initial value problem the data is given by
\begin{equation} \begin{array}{lcl}
e^{-U} & = & cos^2 ( b v \theta(v) ) -
 \frac{\textstyle{a^2} }{\textstyle{b^2}} \, sin^2 ( b v \theta (v) )
 \\[3mm]
e^{-V} & = &
 \frac{ \textstyle{b\, + \, a \, \tan \, ( b v \theta (v) ) } }
      { \textstyle{b\, - \, a \, \tan \, ( b v \theta (v) ) } } \\[1mm]
e^{-M} & = & 1 \;.
\end{array}  \label{ivrosen}
\end{equation}
The limiting values of this result are familiar. If we have just
an impulsive gravitational wave, we must pass to the limit $b \rightarrow 0$
which yields
\begin{equation} \begin{array}{lcl}
e^{-U} & = &1 - a \, v^2 \, \theta (v)   \\
e^{-V} & = &   \frac{ \textstyle{ 1 +  a v \theta (v) }   }
{ \textstyle{1 - a v \theta (v) }  }   \\[1mm]
e^{-M} & = & 1 \;
\end{array}  \label{ivkp}
\end{equation}
as in the case of Khan and Penrose.
Furthermore, in the limit  $a \rightarrow 0$
we have only an electromagnetic shock wave
\begin{equation} \begin{array}{lcl}
e^{-U} & = & 1 - sin^2 ( b v \theta(v) ) \\
e^{-V} & = &  1 \\
e^{-M} & = & 1 \;
\end{array}  \label{ivbs}
\end{equation}
which is the result for the Bell-Szekeres case.

Spacetimes describing colliding plane waves are divided into four
regions:\\[2mm]
\noindent
{\cal Region} \hspace{1mm} I :
$u < 0, v < 0$  empty space before the collision\\
{\cal Region} \hspace{0.5mm} II:  $u > 0, v < 0$  a  plane wave                     \\
{\cal Region} III: $u < 0, v > 0$ another wave travelling in
the opposite direction   \\
{\cal Region} IV:  $u > 0, v > 0$  the interaction region\\[2mm]
\noindent
The initial values given above are on $v = 0$, the boundary between
Regions III, IV and similar results hold on $u = 0$, the boundary between
Regions II, IV, determining the $u$-dependence.
In the latter case the amplitudes of these waves will be
given by different constants, say $ a \rightarrow p $ and
$b \rightarrow q$, {\it cf} eq.(\ref{fg}) in sequel.
The case of Griffiths' solution is a mixture where we have eqs.(\ref{ivkp})
between Regions III, IV and eqs.(\ref{ivbs}) with $u$ replacing $v$
on the boundary between Regions II, IV.

\section{Einstein-Maxwell Equations}

The Einstein-Maxwell field equations governing the interaction
of two plane waves is well-known \cite{sz}.
Starting with the metric (\ref{rosen})
and the Maxwell potential 1-form ${\cal A}$
\begin{equation}
{\cal A} = A d x
\end{equation}
where $A$ depends only on $u, v$, we find a set of Einstein field
equations which
can be grouped into two categories. First we have the
initial value equations
\begin{equation} \begin{array}{lcl}
2 U_{vv} - U_{v}^{\;2} + 2 M_v U_v - V_{v}^{\;2} & = &
     2 \kappa      e^{U - V} A_{v}^{\;2}  \\
2 U_{uu} - U_{u}^{\;2} + 2 M_u U_u - V_{u}^{\;2} & = &
     2 \kappa      e^{U - V} A_{u}^{\;2}  
\end{array}  \label{ivuv}
\end{equation}
and their integrability conditions
\begin{equation} \begin{array}{c}
  U_{uv} - U_{v} U_{u}  =  0  \\
 2 A_{uv} - V_{v} A_{u}  - V_{u} A_{v}  = 0  \\
 2 V_{uv} - U_{v} V_{u}  - U_{u} V_{v} + 2 \kappa e^{U - V} A_u A_v = 0 \\
 2 M_{uv}  +  U_{v} U_{u} - V_{v} V_{u}  + 2 \kappa e^{U - V} A_u A_v = 0
\end{array}  \label{integ}
\end{equation}
where $\kappa$ is Newton's constant in geometrical units.

In eqs.(\ref{integ}) we have the wave equation for $e^{-U}$ and its
solution is immediate from the initial values. The
following two equations are the main equations
and the last equations is irrelevant as
$M$ can be obtained from quadratures once
the main equations are solved.

The problem consists of finding a solution to eqs.(\ref{integ})
satisfying the initial data (\ref{ivrosen}).
Finally, we shall remark that eqs.(\ref{ivkp}), or
(\ref{ivbs}) can be regarded as the solution of an initial value
problem themselves, namely one between Region I and either a
gravitational impulsive wave, or an electromagnetic shock wave
across the null plane $v = 0$ in Region III. In this mini-problem
the first one of eqs.(\ref{ivuv}) serves as the field equation.

\section{Harmonic Maps}

We refer to \cite{es} and \cite{el} for a review and survey of the
principal results on harmonic mappings of Riemannian manifolds.
Here we shall briefly recall the most basic definitions in order
to fix the notation.
We shall consider two Riemannian manifolds endowed with metrics
\begin{equation}       \begin{array}{cc}
  ds^{2} = g_{ik}\,d\,x^i \; d\,x^k  \;, & \;\;\;\;\;\;\; i=1,...,n \\
  ds'^{\;2} = g'_{\alpha \beta}\,d\,y^{\alpha} \; d\,y^{\beta} \;,\;  &
\;\;\;\;\;\;\; \alpha=1,...,n'
\end{array}               \label{mmp}
\end{equation}
and a map
\begin{equation}
f :  {\cal M} \rightarrow {\cal M}'
\end{equation}
between them. This map is called harmonic if it
extremizes the energy fuctional of Eells and Sampson,
$ \delta {\cal I} = 0 $,
\begin{equation}
{\cal I} (f) = \int g'_{\alpha \beta} 
      \frac{\partial f^{\alpha} }{ \partial x^i }
      \frac{\partial f^{\beta} }{ \partial x^k }
g^{ik}   \sqrt{\y \, g \, \y} \, d^n x                      \label{energy}
\end{equation}
where the Lagrangian consists of the trace with respect
to the metric $g$ of the induced
metric $f*g$ on ${\cal M}$. When the target space ${\cal M}'$ is 
1-dimensional, harmonic maps satisfy Laplace's equation on 
the background of ${\cal M}$ and on the other hand
if ${\cal M}$ is 1-dimensional, then
harmonic maps coincide with the geodesics on ${\cal M}'$.
The nonlinear sigma model corresponds to the harmonic map
$f : R^2 \rightarrow S^2$.

There are at least three different ways in which the Einstein-Maxwell
equations (\ref{integ}) can be formulated as a harmonic map.

{\it 1.} We can take a 4-dimensional target space ${\cal M}'$ with the metric
\begin{equation}
  ds'^{\;2} = e^{-U} \; \left( 2\,d\,M \, d\,U \,
+ \, d\,U^2 \, - \, d\,V^2 \,  \right) \;
- \, 2 \kappa \,e^{-V}  \, d\,A^2 \,
\label{4d}
\end{equation}
which has sections of constant curvature and the flat metric
\begin{equation}
  ds^{2} = 2\, d\,u \;d\,v \;
\label{m1}
\end{equation}
on ${\cal M}$. This is the electrovac analogue of the formulation given in
\cite{nh}. It is not the most economical approach because
$M$, which can be obtained by quadratures, appears explicitly
in the metric (\ref{4d}).

{\it 2.} We can get rid of $M$ and consider a
2-dimensional target space with constant curvature
at the expense of regarding $U$ as a given function on ${\cal M}$ which
does not enter into the variational problem as one of the
local components of the harmonic map.
Thus assuming that $e^{-U}$ satisfies the wave equation, as in
eqs.(\ref{integ}), we can take the metric on the target space as
\begin{equation}
  ds'^{\;2} = \ e^{-U} \; \frac{ d{\cal E} \, d{\bar {\cal E}} }
  {\yy Re {\cal E} \yy^2}
\label{2d}
\end{equation}
where
\begin{equation}
 {\cal E} = e^{( V - U )/2} + i \sqrt{\frac{\kappa}{2}} A
\label{ernst}
\end{equation}
is an Ernst potential type of complex coordinate.
The metric (\ref{2d}) is that of a space of constant negative
curvature and ${\cal E}$ is the complex coordinate for the
Poincar\'{e} upper half plane.
There exists another representation, namely Klein's unit
disk for the space of constant negative curvature.
This is obtained by the transformation
\begin{equation}
 {\cal E}   = \frac{ \xi + 1}{ \xi - 1}
\label{klein}
\end{equation}
and
\begin{equation}
  ds'^{\;2} = \, e^{-U} \, \frac{ d \xi \; d {\bar \xi} }
  { ( 1 - \xi {\bar \xi} )^2 }
\label{2dk}
\end{equation}
is the resulting form of the metric. Frequently this
is the most convenient representation.  The original Matzner-Misner
\cite{mm} as well as the Neugebauer-Kramer \cite{nk}
formulations are of this type.

The metric on ${\cal M}$ is the same as the one in eq.(\ref{m1}).

This approach is by far the most common procedure followed in the
literature.

{\it 3.} It is possible to reformulate the reduced problem
by avoiding the {\it ad hoc} introduction of $U$ on ${\cal M'}$
at the expense of adding a new Killing direction
to ${\cal M}$ where the magnitude of the Killing vector is $e^{-U}$.
In this case we have
\begin{equation}
  ds'^{\;2} =  \frac{ d \xi \; d {\bar \xi} }
  { ( 1 - \xi {\bar \xi} )^2 }
\label{2d1}
\end{equation}
where the definition of $\xi$ is the same as before and
\begin{equation}
  ds^{2} = 2\, d\,u \;d\,v \; - \,e^{- 2 U} \, d\,z^2
\label{m3}
\end{equation}
where we consider the mapping to be independent of $z$
and once again $U$ is specified {\it a priori}. It appears that this
possibility has not been discussed before in the literature.

The first formulation where it is not necessary to introduce information
from outside the variational principle is attractive.
However, the close resemblance of the latter formulations
to non-linear $\sigma$-models is an advantage.

\section{Solutions}

The metric (\ref{m1}) on {\cal M} is not in a form most suitable
for the construction of harmonic maps. We need to rewite it
using new coordinates in such a way that
some information about the initial data
(\ref{ivrosen}) is already incorporated into the system
with the result that we can look for simple harmonic maps
automatically satisfying the initial data.
For this purpose we start with the wave equation for $e^{-U}$
and following Szekeres \cite{sz} write its solution as
\begin{equation}
e^{-U} = f(u) + g(v)   \label{ufg}
\end{equation}
where from eq.(\ref{iv}) we know that
\begin{equation} \begin{array}{lcl}
f & = &   \frac{\textstyle{1} }{\textstyle{2}}
- \left( 1 +  \frac{\textstyle{a^2} }{\textstyle{b^2}} \right)
 \, sin^2 ( b v \theta (v) )  \, ,  \\[2mm]
g & = &   \frac{\textstyle{1} }{\textstyle{2}}
- \left( 1 +  \frac{\textstyle{p^2} }{\textstyle{q^2}} \right)
 \, sin^2 ( q v \theta (v) ) \, .
\end{array}            \label{fg}
\end{equation}
We can now consider a trivial coordinate transformation
$$ u \rightarrow f(u) \;\;\;\;\;\;\; v \rightarrow g(v) $$
which amouts to a replacement of $u, v$ by $f, g$ in
the Einstein-Maxwell equations (\ref{integ}).
We shall further introduce the following definitions
which are formally the same  as those given by Khan and Penrose
\begin{equation} \begin{array}{lllllll}
p & = & \sqrt{  \frac{\textstyle{1} }{\textstyle{2}} - f }
& \;\;\;\;\; &
q & = & \sqrt{  \frac{\textstyle{1} }{\textstyle{2}} - g }\\[3mm]
r & = & \sqrt{  \frac{\textstyle{1} }{\textstyle{2}} + f }
& \;\;\;\;\; &
w & = & \sqrt{  \frac{\textstyle{1} }{\textstyle{2}} + g }
\end{array}            \label{pqrw}
\end{equation}
in terms of which it will be convenient to introduce
new coordinates on {\cal M}.

In the vacuum case with non-collinear polarization we had found that
\begin{equation}
\sigma = p w - q r \;\;\;\;\;\;  \tau = p w  + q r
\label{st}
\end{equation}
were useful new coordinates because of two reasons. First of all,
$\sigma$ and $\tau$ can be recognized as prolate spheroidal
coordinates \cite{z} and we have the Kerr-Tomimatsu-Sato solutions
of the main equations. Furthermore, the simplest and the most
familiar solution of this type
\begin{equation}
\xi = cos (\alpha)  \tau + i sin (\alpha) \sigma
\label{kerr}
\end{equation}
satisfies the initial data.

This situation changes for the electrovac case. It turns out that
a useful definition of new coordinates is
\begin{equation}
\sigma = \frac{ p - q }{ r + w } \;\;\;\;\;
        \tau =  \frac{ p + q }{ r + w }
\label{zip}
\end{equation}
in terms of which the metric on {\it M} is given by
\begin{equation}
  ds^{\;2} =
   \frac{\textstyle{d\,\tau^2}}{\textstyle ( 1 + \tau^2 )^2}
   - \frac{\textstyle{d\,\sigma^2}}{\textstyle ( 1 + \sigma^2)^2 }
\label{m2}
\end{equation}
and from eqs.(\ref{ufg}) and (\ref{fg}) we have
\begin{equation} \begin{array}{ll}
e^{-U} & =  \frac{\textstyle {(1 - \sigma^2 ) (1 - \tau^2 ) } }
{\textstyle {(1 + \sigma^2 ) (1 + \tau^2 ) } } \\[3mm]
 & = 1 - p^2 - q^2   .
\end{array}    \label{ust}
\end{equation}
Then the main equations become
\begin{equation}    \begin{array}{cc}
- \frac{ {\textstyle ( 1 + \sigma^2 )^2}}
{{\textstyle ( 1 - \sigma^2 )}}
\, \frac{{\textstyle \partial }}
{{\textstyle \partial \sigma}}
\left(  ( 1 - \sigma^2 )
\frac{{\textstyle \partial \xi}}
{{\textstyle \partial \sigma}}   \right)
+ \frac{{\textstyle ( 1 + \tau^2 )^2}}
{{\textstyle ( 1 - \tau^2 )}}
\, \frac{{\textstyle \partial }}{{\textstyle \partial \tau}}
\left(  ( 1 - \tau^2 )
\frac{{\textstyle \partial \xi}}{{\textstyle \partial \tau}}
\right)        & \\[3mm]
= \frac{{\textstyle  2 {\bar \xi}}}
{{\textstyle  \xi {\bar \xi} - 1 }}       \left[
- ( 1 + \sigma^2 )^2
\left( \frac{{\textstyle \partial \xi}}
{{\textstyle \partial \sigma}} \right)^2
+ ( 1 + \tau^2 )^2
\left( \frac{{\textstyle \partial \xi}}
{{\textstyle \partial \tau}} \right)^2   \right]             &
\end{array}
\label{yeni}
\end{equation}
which is similar to the prolate spheroidal case but differs from it
in some important respects. Its advantage lies in the fact that
\begin{equation}
\xi = \epsilon \sigma, \;\;\;\;\;\;\;  \epsilon^2 = \pm 1
\label{s1}
\end{equation}
is a solution of eq.(\ref{yeni}) that leads to the Bell-Szekeres
solution. Furthermore it can be readily verified that
$\xi = \epsilon \tau $ is also a solution as in eq.(\ref{kerr})
which is again the Bell-Szekeres solution.
In terms of these coordinates the Bell-Szekeres solution is given by
\begin{equation}
  ds^{\;2} =
   \frac{\textstyle{d\,\tau^2}}{\textstyle ( 1 + \tau^2 )^2}
   - \frac{\textstyle{d\,\sigma^2}}{\textstyle ( 1 + \sigma^2)^2 }
- \left(
\frac{\textstyle 1 - \tau^2}{\textstyle 1 + \tau^2 } \right)^2 \, d x^2
- \left(
\frac{\textstyle 1 - \sigma^2}{\textstyle 1 + \sigma^2} \right)^2 \, d y^2
\label{bz2}
\end{equation}
which may help to clarify the meaning of $\sigma$ and $\tau$.

So the remaining problem is to a find a one-parameter complex
solution of eq.(\ref{yeni}) that reduces to eq.(\ref{s1}).
Such a solution will be of physical interest as it will automatically
satisfy the proper initial data.

\section{Conclusion}

The only proper conclusion of a paper such as this, namely
the exact solution of eqs.(\ref{yeni}) satisfying the initial data
in eqs.(\ref{ivrosen}) is missing. In this paper I have given a list
of the essential properties that we must require from a physically
acceptable solution describing the interaction of plane impulsive
gravitational and electromagnetic shock waves and presented some
preparatory material towards such a solution. In my case this
solution has been missing since 1978 and that is why I felt it
inappropriate to publish the work reported here earlier.
However, I now feel that the abundance of
so many irrelevant ``solutions" of this problem in the literature
has made the presentation of the real problem imperative.

\section{Acknowledgement}

This work was in part supported by The Turkish Scientific Research
Council T\"{U}BITAK under tbag-cg-1.

\end{document}